# Voltage control of nuclear spin in ferromagnetic Schottky diodes

R. J. Epstein, J. Stephens, M. Hanson, Y. Chye, A. C. Gossard, P. M. Petroff, and D. D. Awschalom

*Center for Spintronics and Quantum Computation, University of California, Santa Barbara, California 93106*

**Abstract**

We employ optical pump-probe spectroscopy to investigate the voltage dependence of spontaneous electron and nuclear spin polarizations in hybrid MnAs/n-GaAs and Fe/n-GaAs Schottky diodes. Through the hyperfine interaction, nuclear spin polarization that is imprinted by the ferromagnet acts on conduction electron spins as an effective magnetic field. We demonstrate tuning of this nuclear field from <0.05 to 2.4 kG by varying a small bias voltage across the MnAs device. In addition, a connection is observed between the diode turn-on and the onset of imprinted nuclear polarization, while traditional dynamic nuclear polarization exhibits relatively little voltage dependence.

Recent work has revealed that a ferromagnetic layer, such as MnAs or Fe, deposited on an n-type GaAs epilayer can produce large spontaneous electron spin polarization in the GaAs conduction band under optical excitation.[1,2] Depending on the type of ferromagnetic material, these carriers become polarized either parallel or antiparallel to its magnetization. A theoretical explanation of this phenomenon based on spin-dependent scattering at the ferromagnet-semiconductor interface and possible device applications have also been proposed recently.[3] In addition, repetitive optical excitation over several minutes leads to a highly polarized nuclear spin system due to the hyperfine interaction with these electrons. This spin flip-flop process is known as dynamic nuclear polarization (DNP) and has be extensively studied.[4,5,6,7] The nuclear polarization tracks the magnetization of the ferromagnetic layer and acts back on conduction electron spin as an effective magnetic field (up to 14 kG) that strongly influences the coherent electron spin dynamics.[1,2] Direct evidence of the nuclear spin polarization in these systems has been provided by all-optical nuclear magnetic resonance.[1,8,9,10]

Here we investigate the influence of a voltage applied across metallic-ferromagnet/n-GaAs Schottky diodes processed from similar structures. We find that the ferromagnetically imprinted nuclear polarization (termed magnetic DNP henceforth) is strongly voltage dependent, being largest at positive bias while almost completely suppressed under reverse bias. Comparison to the diode electrical characteristics reveals a connection between the current turn-on and onset of magnetic DNP. This is in contrast to traditional optical DNP, i.e. due to an electron spin imbalance generated by polarized optical excitation, which is measured to have relatively weak voltage dependence. Analysis of these results suggests that the observed voltage dependence of magnetic DNP is due to changes in the degree of spontaneous electron spin polarization. The voltage dependence is consistent with spin-dependent electron scattering at the ferromagnet/GaAs interface. However, the sign of spin polarization does not appear to be set by the Schottky barrier height as predicted,[3] but may be due to another factor such as the ferromagnet's spin-dependent density of states.

Two sample series are grown by molecular beam expitaxy with the following structure: ferromagnet/n-GaAs/Al$_{0.75}$Ga$_{0.25}$As(400 nm)/GaAs(001)-substrate. For one series, the ferromagnet is 25 nm of type A MnAs,[11] the n-GaAs is 500 nm thick, and the substrate is semi-insulating. For the other series, the ferromagnet is 10 nm of Fe covered with 5 nm of Al to prevent oxidation,[12] the n-GaAs is 100 nm thick, and the substrate has n+ doping. The ferromagnetic and semiconducting portions of the samples are grown in separate chambers to achieve higher quality GaAs. An As cap is used to protect the GaAs surface during transfer of the sample to the ferromagnetic chamber. The n-type doping (Si: ~7 x 10$^{16}$ cm$^{-3}$) of the GaAs epilayer is selected as a compromise between long electron spin lifetime[13] and ease of Ohmic contacting. For electrical measurements, a portion of the MnAs is etched away and both the GaAs and MnAs are contacted with In. The GaAs contacts are annealed prior to contacting the MnAs. For the Fe samples, the thinner n-GaAs epilayer is fully depleted so the n+ substrate is contacted instead. For optical transmission measurements, samples are mounted on fused silica and a region of the GaAs substrate is removed by a chemically selective etch. The left inset of Fig. 1 shows a schematic of the device structure.

Electron spin dynamics in these samples are probed with time-resolved Faraday rotation (TRFR)[14] using a mode-locked Ti:Sapphire laser that produces 100 fs pulses at a repetition rate of 76 MHz. The pulse train is tuned to an energy near the GaAs band gap (1.52 eV at T = 5 K) and is split into circularly polarized pump and linearly polarized probe beams that have a power ratio of 20:1 and are focused to overlapping ~50 μm spots on the sample. Due to the optical selection rules in GaAs, a given pump pulse excites carriers with net spin aligned along the pump path.[7] After a time Δt, the corresponding probe pulse is transmitted through the sample with its linear polarization axis rotated by an angle $\theta_F$ proportional to the net spin along the probe path. In order to measure coherent spin dynamics, an external field $\mathbf{B}_{app}$ is applied transverse to the pump beam, inducing coherent spin precession in a plane normal to $\mathbf{B}_{app}$. This precession leads to a periodic oscillation of $\theta_F$ vs Δt, where Δt is varied with a mechanical delay stage. As holes rapidly spin-relax in bulk GaAs due to spin mixing in the valence band,[15] the evolution of $\theta_F$ is dominated by electron spin precession and is generally described by $\theta_F = Ae^{(-\Delta t/T_2^*)}\cos(\omega_L t + \phi)$, where A is the initial Faraday rotation angle, $T_2^*$ is the effective transverse spin lifetime, $\omega_L$ is the Larmor precession frequency and $\phi$ is the phase. Here, $\omega_L = g\mu_B B_{tot}/\hbar$, where g is the Landé g factor, $\mu_B$ is the Bohr magneton and $\mathbf{B}_{tot}$ is the total field—the sum of $\mathbf{B}_{app}$ and the nuclear field $\mathbf{B}_n$. The latter is given by $\mathbf{B}_n = A\langle\mathbf{I}\rangle/g\mu_B$, where A is the contact hyperfine constant and $\langle\mathbf{I}\rangle$ is the average nuclear spin.[7] Thus, the electron Larmor frequency, obtained through TRFR, is used as a magnetometer to probe the nuclear spin system.[8,9,10,16]

Figure 1 shows TRFR Δt scans at three representative bias voltages under excitation normal to the MnAs sample surface with $B_{app}$ = 2.5 kG and T = 5 K. The magnetic field is applied in-plane along the MnAs easy axis and is somewhat larger than the coercive field (~1 kG). In addition to electron spin polarization generated by traditional optical pumping, spontaneous spin polarization is produced along the magnetization,[2] which is does not contribute to Faraday rotation directly but is detected by changes in $\omega_L$ due to dynamic polarization of the nuclear spin system (magnetic DNP). At a bias of 0 V, $\omega_L$ is slowest and $T_2^*$ is longer than the delay scan range. As the voltage is progressively increased toward positive bias, the precession frequency increases while the spin lifetime shortens. After setting a new voltage value, the TRFR signal equilibrates over several minutes due to nuclear spin relaxation. The reduction in $T_2^*$ may be explained by a non-uniform nuclear polarization profile caused by the carrier spin distribution and resulting in inhomogeneous dephasing of the electron spin ensemble.[17] Furthermore, fits to the data yield that $\omega_L/2\pi$ is tuned over a 1.3 GHz range with a 1 V change in bias—an initial indication of voltage-tunable nuclear spin polarization in these devices.

The inset of Fig. 1 shows similar TRFR data for the Fe sample with a coercive field ~50 G and $B_{app}$ = 1 kG. Spin precession is fastest at positive bias, and slows as the bias is decreased to negative values. In previous work on Fe/GaAs,[2] the nuclear field was found to be antiparallel to $\mathbf{B}_{app}$ under similar conditions. Thus, it is possible to have $\omega_L \approx$ 0 as shown for –1.7 V, indicating that $\mathbf{B}_n \approx -\mathbf{B}_{app}$. In this device, $B_{tot}$ is tunable from 0 ± 0.4 kG (accuracy limited by $T_2^*$) to 15 kG, corresponding to ~9 GHz variation in precession frequency for a 3.2 V change in voltage. Based on a calculation[18] for bulk GaAs, a field of 14 kG corresponds to a nuclear spin polarization of ~26%. While the tunable range of nuclear polarization is larger for the Fe device, biasing through the AlGaAs layer modifies the current-voltage curve by impeding current flow at low voltages making the following comparisons more difficult. Therefore, we focus on the MnAs device, in which the n-GaAs epilayer is contacted directly, because the diode electrical characteristics more clearly reflect the band bending near the MnAs/GaAs interface.

Current-voltage curves for the MnAs device are plotted in Fig. 2 for pump powers of 2.0 and 0.8 mW and with the laser blocked. The latter two curves are very similar apart from a vertical shift due to photocurrent contributions. The 2 mW curve, however, exhibits a more gradual turn-on of the diode, which may be explained by carriers partially screening the Schottky barrier.

To compare these electrical characteristics to the response of the nuclear field, TRFR $\Delta t$ scans were taken at successive voltages over the same range for both pump powers and with $B_{app}$ = 2.5 kG. The TRFR data in Fig. 1 are excerpts from the 2.0 mW dataset. Fits to the TRFR data yield $\omega_L$ from which $B_{tot}$ is calculated as shown in Fig. 2. A 10-minute pause after each voltage step allowed the nuclear polarization to reach its steady-state value. To suppress optical DNP, the time-averaged optically oriented electron spin was canceled out by switching the pump from right to left circular polarization (RCP to LCP) at 50 kHz with a photoelastic modulator. Upon completing the entire scan, TRFR data was retaken at +0.5V after a 30-minute pause (star symbol). The small deviation (1.6%) between the two data at +0.5 V verifies that the pause between voltage steps is sufficiently long to accurately reflect changes in nuclear polarization.

For both excitation powers, $B_n$ is largest at positive bias and saturates at high bias, whereas it is almost completely quenched at negative bias ($B_{tot} \approx B_{app}$). The residual $B_n$ (~50 G) may be due to imperfect suppression of optical DNP. The lower pump power yields less nuclear polarization in accordance with previous results[1] and as expected for DNP.[7] Moreover, the turn-on of the diode and the onset of nuclear polarization appear to be connected—both occur at similar voltages and shift to lower voltage as the pump power is increased. Note that the g factor ($|g| = 0.42$) used to calculate $B_{tot}$ is obtained using a pump power of ~80 µW, where no nuclear polarization is detected and $\omega_L$ is constant over the measured voltage range (not shown). In addition, DNP through electrical spin injection[19] without photoexcitation is not detected, which may be due to the large area of the ferromagnetic contact (~2 mm$^2$) resulting in low current densities.

To confirm that possible changes in the g factor are not responsible for the observed voltage dependences, TRFR data was taken at T = 180 K where there is negligible nuclear polarization (open squares).[1] In this case $B_{app}$ = 5 kG and, therefore, $B_{tot}/2$ is plotted for comparison. Using $g = \hbar\omega_L/\mu_B B_{app}$, one finds that the g factor changes by <1% over the voltage range displayed.

In order to gain further insight into the origin of this bias-dependent nuclear polarization, we also measure optical DNP in the same devices. This is achieved using a static wave-plate to circularly polarize the pump and by rotating the sample 20° as in Fig. 3 inset. Due to optical refraction, there is then a non-zero component of electron spin $\mathbf{S}_\parallel$ injected parallel to $\mathbf{B}_{app}$. These conditions lead to a large nuclear polarization in addition to that produced by the ferromagnetic layer.[20] In the upper panel of Fig. 3, $B_{tot}$ vs bias is plotted for LCP and RCP pumps with $B_{app}$ = 5 kG and 0.8 mW pump power.[21] For RCP, comparison between voltage sweeps from 0 V upward (solid squares) and from 1.1 V downward (open squares) shows that the error in $B_{tot}$ due to the nuclear time-constant is <9%. Furthermore, the data indicate that optical DNP is relatively insensitive to voltage bias. This observation can be made quantitative by separating the optical and magnetic DNP contributions. As RCP and LCP pulses inject electron spin with opposite directions, **B**$_n$ also has opposite sign and adds to or subtracts from **B**$_{app}$. Assuming that the total field is given by B$_L$ = B$_{mag}$ + B$_{opt}$ + B$_{app}$ and B$_R$ = B$_{mag}$ − B$_{opt}$ + B$_{app}$, for LCP and RCP pumps respectively, one can obtain B$_{mag}$ = (B$_L$ + B$_R$)/2 − B$_{app}$ and B$_{opt}$ = (B$_L$ − B$_R$)/2, the nuclear fields due to magnetic and optical DNP, respectively. The lower panel of Fig. 3 shows these calculated values, in addition to data (solid circles) reproduced from Fig. 2. Note that B$_{mag}$ is scaled by a factor of two for the following comparisons. While B$_{opt}$ (open circles) varies by ~3% over the entire voltage range, B$_{mag}$ (solid squares) has very similar voltage dependence to that of Fig. 2 despite the scale factor and a small vertical shift. The reduced magnitude results from replacing the photoelastic modulator with a mechanical chopper in the pump path for lock-in detection, which halves the average pump power. The vertical shift is likely due to imperfect laser polarization giving slightly different |B$_{opt}$| for RCP and LCP pumps.

Insight into the dramatic voltage-response of magnetic DNP is provided by the weaker response of optical DNP. Previous work[1,2,7] suggests that both types of DNP are consistent with $\mathbf{B}_n \sim C\mathbf{B}_{app}(\langle\overline{\mathbf{S}}\rangle \cdot \mathbf{B}_{app})/|\mathbf{B}_{app}|^2$, where $\langle\overline{\mathbf{S}}\rangle$ is the time-averaged electron spin, and $C$ accounts for electron-nuclear wave function overlap and electron spin "leakage" through relaxation processes other than hyperfine-induced flipping of nuclear spin.[7] The optical DNP data shows that $C$ is nearly constant as a function of voltage, which implies that changes in $\langle\overline{\mathbf{S}}\rangle$ may be responsible for the observed dependence of magnetic DNP. Moreover, the above relation shows that B$_n$ can be used to measure the electron spin component parallel to **B**$_{app}$, which is not probed directly with TRFR in the Voigt geometry used here. While both the initial spin polarization and longitudinal lifetime $T_1$ determine $\langle \overline{S} \rangle$, the measured voltage dependence of $T_2^*$ gives an indication of the electron spin lifetime in the GaAs eplayer and cannot explain the dependence of $B_n$. On the other hand, voltage- and spin-dependent carrier reflection at the GaAs/MnAs interface is consistent with the data. At positive bias, the conduction band is flattened and electrons can scatter off of the interface, whereas as negative bias, the electrons are swept away from the interface. The reported theory focuses on tunneling through the Schottky barrier and predicts a sign change in the electron spin polarization as a function of barrier height.[3] Assuming that the MnAs/GaAs interface is nearly Ohmic, as seen for (111) GaAs,[22] and the barrier for Fe is ~0.7 eV, the theory explains the opposite spin polarization observed for the two ferromagnets.[2] Here, however, a significant Schottky barrier is present for both Fe and MnAs devices, and a sign change as a function of bias voltage is not observed in either device. Thus, another factor may be dominant, such as the spin polarization of the ferromagnetic density of states at the Fermi level, which can have opposite signs in these two materials.[23]

In conclusion, we have demonstrated voltage-tunable nuclear spin polarization and electron spin coherence in ferromagnetic Schottky diodes. Combined with lithographic patterning,[24] these results provide a new means of local spin manipulation for semiconductor spintronics. We acknowledge support from DARPA/ONR N00014-99-1-1096, AFOSR F49620-02-10036, ARO DAAD19-01-1-0541 and NSF DMR-0071888.

**Figure Captions**

FIG. 1. Time-resolved Faraday rotation (TRFR) for MnAs device at specified bias voltages with $B_{app}$ = 2.5 kG and T = 5 K. Data have been vertically offset for clarity. Left inset: device schematic. Right inset: TRFR for Fe device at three bias voltages with $B_{app}$ = 1 kG and T = 5 K.

FIG. 2. Current–voltage curves for specified pump powers (solid lines). $B_{tot}$ vs bias voltage for 2.0 mW (open circles) and 0.8 mW (filled squares) pump powers with $B_{app}$ = 2.5 kG. Also shown is $B_{tot}$ vs voltage for a 2.0 mW pump at T = 180 K and $B_{app}$ = 5 kG (open squares). For the latter curve only, $B_{tot}$ is divided by a factor of 2 for comparison. Inset: $B_{tot}$ vs bias voltage for Fe device with $B_{app}$ = 1 kG and T = 5 K. All sweep directions are from negative to positive voltage except for the inset.

FIG. 3. Upper panel: $B_{tot}$ vs bias voltage for 0.8 mW LCP and RCP pumps with $B_{app}$ = 5 kG. Filled symbols are negative to positive voltage sweeps and open squares are in the reverse direction. Inset: measurement geometry showing non-zero component of electron spin ($S_{\parallel}$) parallel to $\mathbf{B}_{app}$. Lower panel: $B_{opt}$ (open circles) and $B_{mag}$ (solid squares) vs bias voltage obtained from data in upper panel. $B_{mag}$ values are multiplied by 2 and the 0.8 mW data from Fig. 2 is shown for comparison (solid circles).

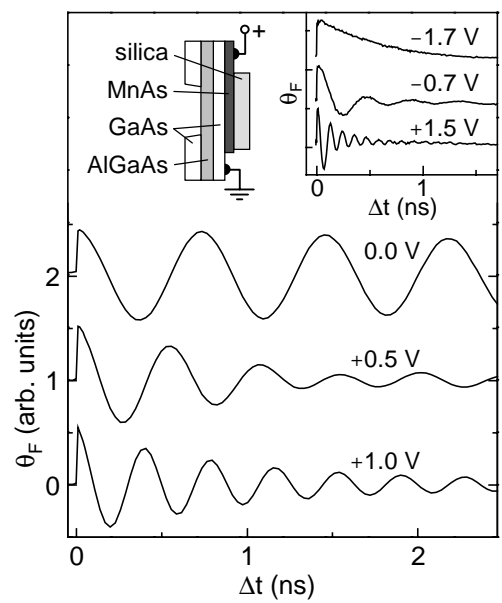

Figure 1. Epstein et al.

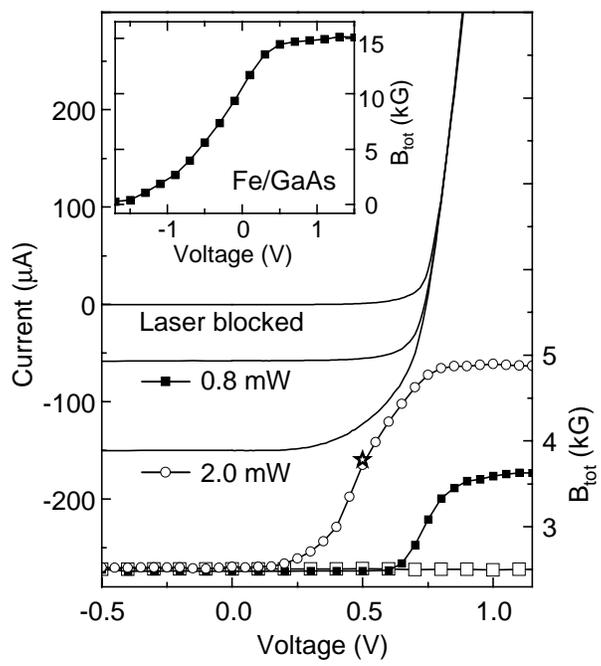

Figure 2. Epstein et al.

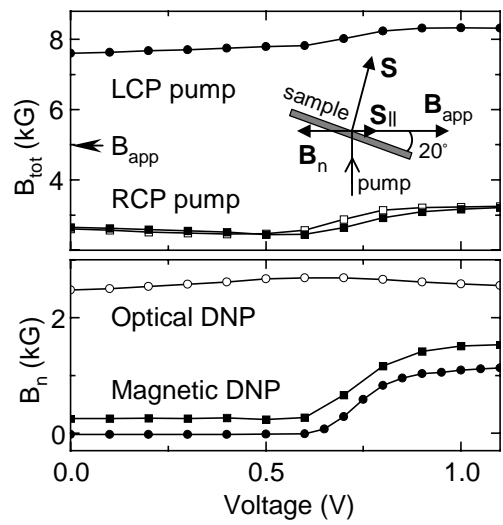

Figure 3. Epstein et al.